\documentclass[letterpaper]{jpconf}
\usepackage{amsmath}
\usepackage{amssymb}
\usepackage{amsfonts}
\usepackage{dsfont}
\usepackage{multirow}
\usepackage{graphicx}
\usepackage{mathrsfs}
\usepackage{graphicx}
\usepackage{stmaryrd}
\usepackage{indentfirst}
\begin{document}
\title{Modular structures and\\ extended-modular-group-structures\\ after Hecke pairs}
\vspace{-16pt}
\author{Orchidea Maria Lecian}
\vspace{-4pt}
\address{Comenius University in Bratislava,
Faculty of Mathematics, Physics and Informatics,
Department of Theoretical Physics and Physics Education- KTFDF,\\
Mlynsk\'a Dolina F2, 842 48, Bratislava, Slovakia;\\
Sapienza University of Rome,
Faculty of Civil and Industrial Engineering,\\ DICEA- Department of Civil, Constructional and Environmental Engineering,\\ Via Eudossiana, 18- 00184 Rome, Italy.
}
\vspace{-4pt}
\ead{orchideamaria.lecian@uniroma1.it; lecian@icra.it}
\vspace{-17pt}
\begin{abstract} The simplices and the complexes arsing form the grading of the fundamental (desymmetrized) domain of arithmetical groups and non-arithmetical groups, as well as their extended (symmetrized) ones are described also for oriented manifolds in $dim>2$. The conditions for the definition of fibers are summarized after Hamiltonian analysis, the latters can in some cases be reduced to those for sections for graded groups, such as the Picard groups and the Vinberg group.The cases for which modular structures rather than modular-group-structure measures can be analyzed for non-arithmetic groups, i.e. also in the cases for which Gelfand triples
(rigged spaces) have to be substituted by Hecke couples, as, for Hecke groups, the existence of intertwining operators after the calculation of the second commutator within the Haar measures for the operators of the correspondingly-generated C$*$ algebras is straightforward.\\ 
 The results hold also for (also non-abstract) groups with measures on (manifold) boundaries.\\
The Poincar\'e invariance of the representation of Wigner-Bargmann (spin $1/2$) particles is analyzed within the Fock-space interaction representation.\\
The well-posed-ness of initial conditions and boundary ones for the connected (families of) equations is discussed. As an example, Picard-related equations can be classified according to the genus of the modular curve(s) attached to the solutions(s).\\
From the Hamiltonian analysis, further results in the contraction of the congruence (extended sub-)groups for non-arithmetical groups for the construction of tori
 is provided as an alternative to the free diffeomorphism group.\\
In addition, the presence of Poincar\'e complexes is found compatible with non-local interactions, i.e. both lattices interactions or spin-like ones.
\end{abstract}
\vspace{-46pt}
\section{Introduction}
The investigations is aimed at understanding what and if are the hypotheses intertwining operators for the second commutator of the  Hecke groups to obtain a von Neumann algebras
by studying the orientation-preserving homotopy groups for closed oriented domains on manifolds (i.e. also the domains on which the group are defined) within the functorial equivalence classes 'borrowed' from the (also, quantum) grouppoid formalism\footnote{
Given a holomic module sheaf of a differential operator on a complex manifold $Q$,\cite{ref03}, the support of the sheaf is a closed homogeneous involution subvariety of the bundle of the complex manifold, on which for the sheaf $\exists$ an exact functorial type form from the holomic module sheaf of a differential operator of finitely properly-chosen generated U-modules. 
Be therefore ${\bf U}=U(\mathfrak{g})$ be the enveloping algebra of $\mathfrak{g}$, endowed with center and at least one ideal \cite{ref03}.It
is possible to prepare exact sequences of injective applications (from the manifold to the fibre) of the composition
of ${\bf U}(\mathfrak{g}$ for the direct product of holomic modules for direct product of the composition of the to the category
generated by the ${\bf U}(\mathfrak{g}$  on the complex manifold according to their
weight. The disjoint union of such injective applications (on appropriately-chosen enclosures) are on a the complex manifold, quotient of a semisimple Lie group and a semisimple Lie algebra \cite{ref03}.\\
Indeed, the proper covariant representations of a pair generate normal *-representations of an enveloping von Neumann algebra compares the covariance C*-algebra of the pair \cite{ref02}.}.
\vspace{-13pt} 
\section{Geometrical definitions}
The algebraic structures for the generators of the subgroups are examined not necessarily as modular structures, s.t. comparison with the Gelfand classes (for identifying (also Hecke) degenerate spectrum sequences of the multi-variable operators) is possible.\\  
The homotopy invariance(s) (classes) for oriented manifold are evaluated after this construction \cite{ref01}.\footnote{ 
These domains can be arbitrarily extended with the restrictions of \cite{ref03} imposed on \cite{ref35},
 and compared by means of the free (known) Hamiltonian flow for
the so-defined algebraic structures (whose eigenvalues have been nevertheless at least conjectured) identifying the homotopy invariance(s classes) of the second commutator
for the considered Hecke (sub-) group. The definition(s) for the fiber in any number of dimensions and also in the presence of Poincar\'e complexes allows to extend the analysis also for non-arithmetical groups.}.
The conditions for the possibility to invert the maps of two unital algebra $(A,B)$ to define 
 the homotopy equivalence classes about the manifolds they identified after the definition of the intertwiners.
\vspace{-14pt}
\subsection{Geometrical study: algebroids}
Let $A$ and $B$ be {\it unital algebras}. {\bfseries {\slshape {\bfseries {\slshape Def.}}}}: a Hopf algebroid is defined by \cite{ref1} a structure on $A$, over $B$, consisting of $i)$ the source map: algebra homomorphism $\alpha : B \rightarrow A$; $ii)$
the target map: algebra anti-homomorphism $\beta : B \rightarrow B$ satisfying $\alpha(a)\beta(b) = \beta(b)\alpha(a)$, for all $a, b \in A$. $\boxempty$ \\
Let $A$ and $B$ be {\it differential graded algebras}. {\bfseries {\slshape Def.}}: a \textbf{Hopf bialgebroid}\normalsize $\ \ $ is defined by a structure on
$A$, over $B$, which admits an algebra anti-isomorphism $\varsigma : A \rightarrow A$, where for which all the needed maps must both
$i)$ be compatible with the differentials; $ii)$ be of degree $0$.\\
The map$\rho$ from the {\it Hopf cyclic cohomology of} $H(G)$ s.t. $H(G)$ is equal to the cyclic cohomology of the algebroid (i.e. the double complex of sheaves of differential forms on the simplicial manifold) to the grouppoid sheaf cohomology
of the sheaf $C$ on the fiber either is the identity, or is an isomorphism. $\boxempty$\\ 
The proof for the inverse map $\rho^{-1}$ is given in\cite{ref2}.\vspace{4pt}\\
The $K$ theory of the corresponding $C*$ algebra of operators is defined at least on a subgroup of the (manifold comprehending the) invariant(s) of these transformations which defines the space spanned by the maps, the antipode map and the co-unit map. It is based on its {\it double affine degenerate Hecke algebra}, and can be extended to the {\textbf non-degenerate}, i.e. to the {\textbf q-deformed} case.\vspace{4pt}\\
{\bf Looking for the solution on the fiber}\\
Given a KW section,\\
$\bullet$  the map $\varrho$ between the composition of the action of $B$ on Lie Groups on $H$
by the K-theory of the reduced C* algebra of operation on Lie groups
is an isomorphism;\\
$\bullet$ the map $\tilde{\iota}$ between the C*algebra of the action of $\Gamma$ on Lie Groups on $H$and that on $G$
is an isomorphism, i.e. after the definitions of gradings for Vinberg groups induced by the same homomorphism on cyclic cohomology for which:\\
${}_{\ \ \ \ } \bullet$ $\exists N\in {\mathbb N}$ s.t. the $N-th$ iteration of $\alpha$, $\alpha^{n_1}$ is trivial (idempotent) $\forall n_i<N$ independently of $i$,\\
${}_{\ \ \ \ } \bullet$  $\exists$ an assembly injective map(s) from idempotent transformations on gruppoid transformations which is an isomorphism,\\
${}_{\ \ \ \ } \bullet$ $\exists$ and injective map  from the fibration of these transformations to these transformations themselves. \vspace{4pt}\\
The {intertwining operators for the mentioned C* algebras} (i.e. \itshape{double affine Hecke algebras}\upshape $\ \ $) are the Dunkl operators $D_j$ for $\beta$ non-negative integer s.t. $i)$ $D_j$ preserve the space of polynomials of variables $x_1, ..., x_N$;
$ii)$  $D^A_j=\frac{\partial}{\partial x_j}+\beta\sum_{k\neq j, j=1}^{j=N}\frac{1-s_{jk}}{x_j-x_k}$, with $s_{ij}$; $ii_a)$ elements of the symmetric group $S({\mathbb N})$; $ii_b)$ act on functions of $x_1$,..., $x_N$ as operators which permute the arguments $xi$ and $xj$. $\boxempty$ \vspace{4pt}\\
{\it Properties of quantum grupoids}
Quantum gruppoids are based on the degenerate double affine Hecke algebra \footnote{a {\it degenerate double affine Hecke algebra} is {\bfseries {\slshape Def.}}: the \textbf{triple} $( \left\{ s_{ij}, x^{\pm}_j, \tilde{D}^A_j \right\} )$ with $s_j\equiv s_{j,j+1}$ $(j = 1,..., n − 1)$ (simple) translations, (isomorphic to the) {\textbf degenerate affine Hecke algebra}.
{\it Intertwiners} between two representations decompose as direct products on different ( among each set and among the two sets- orthogonal) polynomials for mapping on algebra representations after  $sl(2)$ -induced algebra automorphisms. $\boxempty$\\
The map $\sigma$, the structure form $A$ to $\sigma$, the algebra $B$ are applicable to non-symmetric case. As a consequence, it is possible to construct raising operators and shift operators for Jack polynomials
 $P^\alpha_\lambda$, where  $P^\alpha_\lambda=<p_\lambda, p_\mu>_\alpha=\alpha^{l(\lambda)}z_\lambda\delta_{\lambda\mu}$.$\boxempty$ \\
Let $q$ be a root of unity of order $d>1$, with $d$ odd integer; let $A$ be a Hopf algebra
with two generators, $E$, $K$,
 connected with $\theta$, $\epsilon$, s.t. $S(\kappa) = \kappa^{−1}$, $S(\tilde{\eta}) = - \kappa^{−1}\tilde{\eta}$, as in \cite{ref5}. The $sl(2)$ \itshape{induced algebra automorphism}\upshape $\ \ $ is therefore $\theta = q^2S^{−2}$, $\theta(\kappa) = q^2 \kappa$, $\theta(\tilde{\eta}) = \tilde{\eta}$ as in \cite{ref6}, where $S(K) = K^{−1}$, $S(E) = − EK^{−1}$, $\epsilon(K) = 1$, $\epsilon(E) = 0$.}. 
It is expected that the results given here extend to the \textbf{non-degenerate} case, i.e. to the {\textbf q-deformed case}.

 
%
%

\begin{figure}[htbp]
\begin{minipage}{15pc}
\includegraphics[width=15pc]{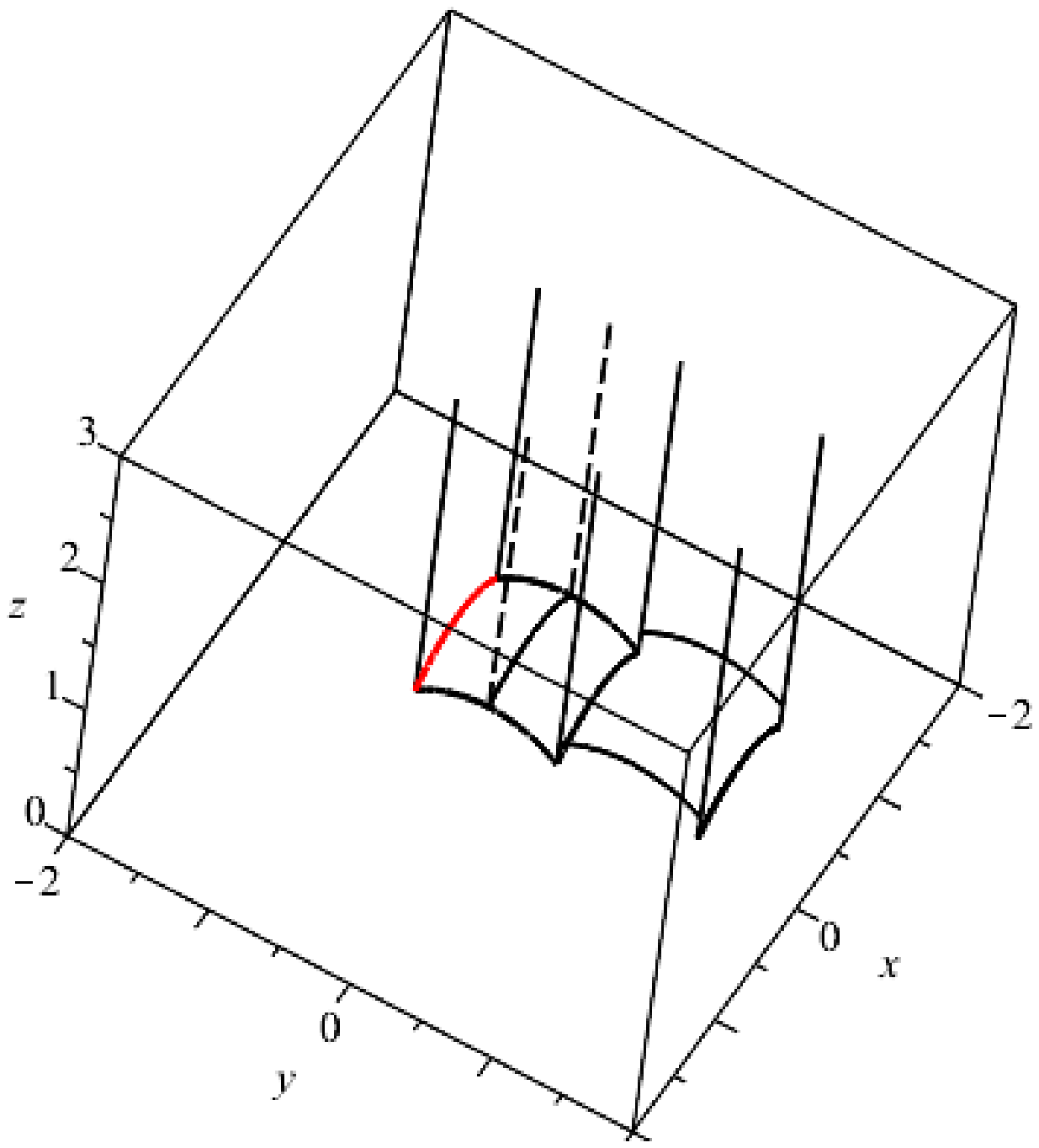}
\caption{\label{labelfigure5} \small $n=1$ { \bf Grading for the  subgrouppal structure}    $\Gamma_0(Pic)$. \scriptsize The domain of the subgrouppal structure for the $n=1$ grading of the subgrouppal structure $\Gamma_0(Pic)$;\\
the simplex $u_1=0\pm\frac{m'}{2}$ is sketched, $m'=2m+1$. \normalsize}
\end{minipage}\hspace{2pc}%
\begin{minipage}{15pc}
\includegraphics[width=15pc]{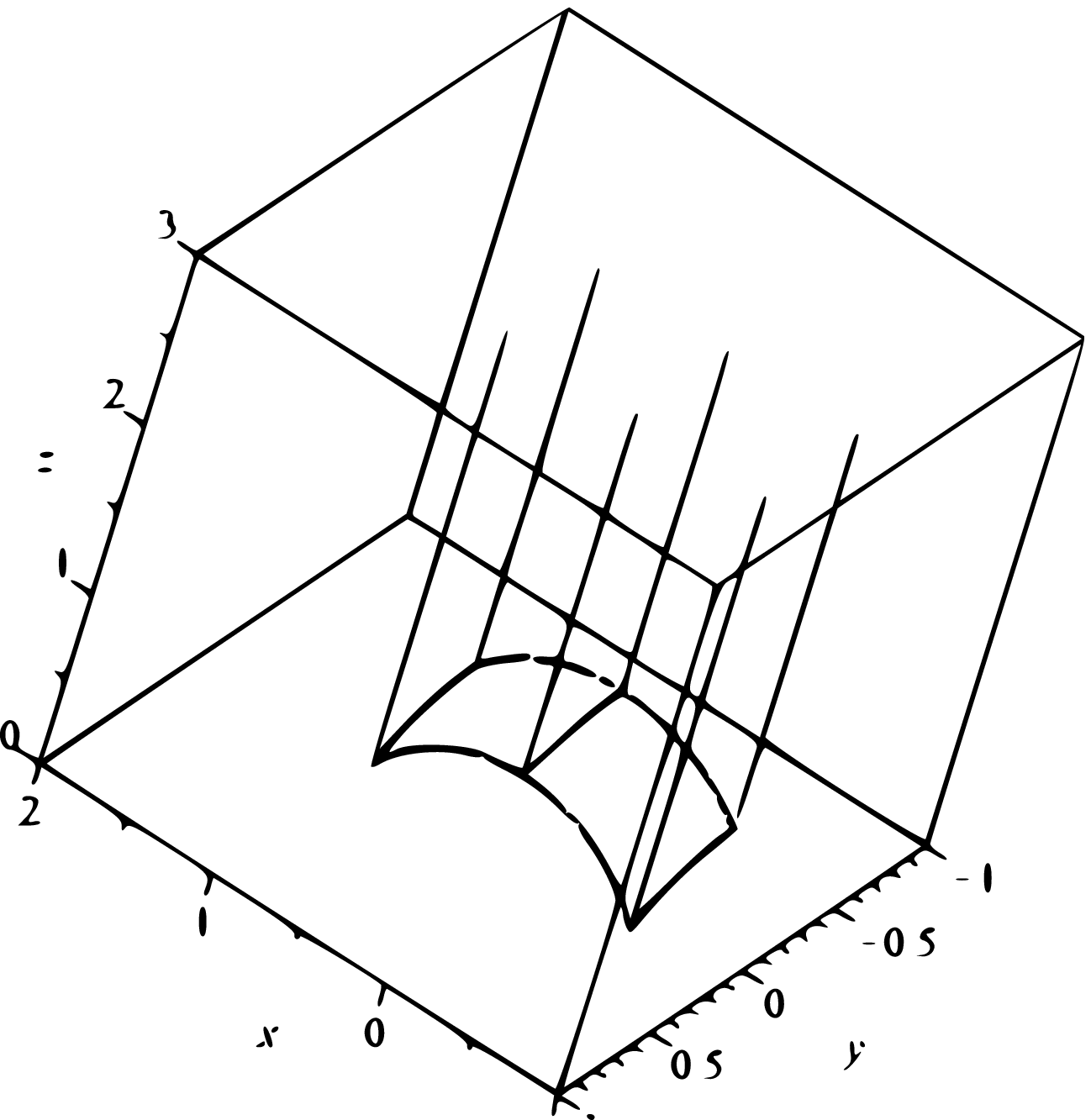}
\caption{\label{labelfigure6} {\small \bf The subgrouppal structure}   $\ \ $$\Gamma_0\left(x_\beta=\frac{\sqrt{5}}{3}\right)$ $\ \ $ {\bf of the Vinberg group}. \scriptsize The domain of the  subgrouppal structure for the Vinberg (non-arithmetical) group characterized by\\
$\bullet \ \ $ an angle $\beta=\frac{4}{5}\frac{\pi}{2}$, i.e. $x_\beta\simeq \frac{\sqrt{5}}{3}$;\\
$\bullet \ \ $$-\frac{\sqrt{5}}{3}\leq x\left(\Gamma_0\left(x_\beta=\frac{\sqrt{5}}{3}\right)\right)\leq \frac{\sqrt{5}}{3}$, $0\leq y\leq \frac{1}{2}$;\\
the simplex at $x=\tilde{m}'\frac{\sqrt{5}}{3}$, $\tilde{m}'\in\mathbb{Z}$ is sketched. \normalsize}
\end{minipage} 
\end{figure}
\vspace{-9pt}
\small
\subsection{Geometrical study: after decomposable groups- Extended Hecke groups, Hecke groups, Picard groups and commutator subgroups}
{\it Extended Hecke groups, Hecke groups, Picard groups and commutator subgroups} {\bfseries {\slshape Def.}}: are subgroups realized by a correspondence that arises \cite{ref10} between
 a subgroup of a group of symmetries which defines the tiling of the subdomains, and
 the transitivity (under the (sub-group)pal) action) of the tiling.\vspace{4pt}\\ 
{\it Hecke groups} $H(\lambda)$ $\in$ $PSL(2;R)$ {\bfseries {\slshape Def.}}: are
 orientation-preserving groups, whose generators are $T(z)=-1/z$, $U(z) = z + \lambda$, s.t. $S\equiv TU$, \small $\Rightarrow S(z)=-\frac{1}{z+\lambda}$. A Hecke group $H(\lambda)$ is \itshape{discrete}\upshape $\ \ $ iff $\lambda=\lambda_q=2\cos(\frac{\pi}{q})$, or
{\bfseries {\slshape Def.}}: a \textit{discrete Hecke pair}
$(G,H)$ is given, when $G$ is a discrete group, and $H$ a subgroup.\vspace{4pt}\\
All Hecke groups are subgroups of $PSL(2;Z), \ \ Z\in Z[\lambda_q]$ $\boxempty$.
\vspace{-16pt}
\subsubsection{$\theta$ groups}. Let $\mathcal{G}$ be a reflection group over the $n$-dim Hermitian space $V$, defined on
$\mathcal{R}$ reflections: given $k:\mathcal{G}\rightarrow\mathbb C$, $\exists$ a {\bfseries {\slshape Def.}}: $\mathcal{G}$-invariant parameter family functions, with $\mathcal{H}(\mathcal{G},k)$ the rational Cherednik algebra,
isomorphic to $\mathcal{P}\otimes\mathbb C\mathcal{G}\otimes \mathcal{S}$
\cite{ref11}. $\boxempty$ \\
 The corresponding Hecke-Drinfeld algebra $HD$ is defined as $HD\equiv\mathcal{P}\otimes\mathbb C\mathcal{G}$ \cite{ref12}. \vspace{4pt}\\
The evaluation of the second commutant \cite{ref15}, \cite{ref14} (after the evaluation of the first commutator group \cite{ref13})
 coincides with the results for the extended modular group; for groups generated by three reflections, \cite{ref16}, it is possible compare the congruence subgroups  of the $PGL(2,{\mathbb Z})$ group (a.k.a. \textit{the extended modular group}) and also, in particular, $\Gamma_2\left(PGL(2, {\mathbb Z})\right)$ group, also in the cases \cite{ref16} \cite{ref18} \normalsize wrt\footnote{{\it Kurosh theorem}
 \cite{ref8} for the product of free groups can be compared as a generalization of \cite{ref9}. Be $G$ a group with associative multiplication law, and $g$ one of its elements,
$g\in G$, s.t.
 $i)$ $g=h_1h_2...h^n$, representation of products of two elements $g, g'\in G$, $g=h_1h_2...h^n$, $g'=h'_1h'_2...h'^m$;
$ii)$ $gg'=h_1h_2...h^nh'_1h'_2...h'^m$ invertible associative multiplication law;
$iii)$ $g^{-1}=h_1^{-1}h_2^{-1}...h_n^{-1}$ from the explicit representation $n=$ $\mathfrak{l}$ $(g)$;
\itshape{but}\upshape $\ \ $
$h'_{k+1}\neq h^{-1}_{n-k}$ $\forall$ $0\ge k\ge min(n, m)$: $\Rightarrow$ $\exists$ $!$ the (injective and surjective) \itshape{bijective relation}\upshape $\ \ $$\ \ $ with $\ \ $ $h_{n-k}h'_{k+1}$ of different elements for each component. $\boxempty$ \\
The {\it unit element} $\hat{1}$ {\bf does not contribute to the associative law}. For $F$ a subgroup of $G$, $F\subset G$, {\bfseries {\slshape Def.}}: for an $F\equiv F_1\cap F_2$, $F$ is called the \textbf{mean sub-group} for the sub-groups $F_1$ and $F_2$.
The use of decomposable groups is aimed at trying to avoid the need for truncated quantum group algebra(s) \cite{ref85}. $\boxempty$ \vspace{4pt}\\
{\it Hypergroups}  $\tilde{K}$ are defined as {\bfseries {\slshape Def.}}: $\tilde{K} = \bar{K}*K$, with $\bar{K}$ any copy of $K$.\\
The corresponding Hecke algebra is the $*$-algebra $M(\tilde{K})$ of bounded measures
on $\tilde{K}$.\vspace{4pt}\\
Let $M(\tilde{K})$ be endowed with a $\delta$ Dirac distributional measure (on a Hilbert space $\mathcal{H}$) \cite{ref7}.\\
The definition of {\it Gelfand pairs for Hecke groups} also extends given $M(\tilde{K})$ endowed with a $\delta$ Dirac distributional measure (on a Hilbert space $\mathcal{H}$) \cite{ref7}. {\bfseries {\slshape Def.}}: $(G, K)$ is a \textit{Gelfand pair} iff
 $G$ locally compact group (endowed with neutral element, where the latter is not defined by idempotence under iterations and the Haar measure on $G$ is unimodular), 
$K$ a compact subgroup. $\boxempty$ \\ 
{\bfseries {\slshape Def.}}: $(G,K)$ is a \textit{generalized discrete Hecke pair} iff
 $\nu$ a non-degenerate representation of $c_0(K)$ on a (Hilbert space H), and $V$ be a $*$-representation
of $M(\tilde{K}$). Furthermore, {\bfseries {\slshape Def.}}: if $\exists$  intertwiners $\nu 1\otimes M$ $\Rightarrow V$ is a representation of $M(\tilde{K})$ such that $(\nu, V)$ is a \textit{covariant pair} $\Leftrightarrow$ $(\nu, V )$ is a covariant pair iff
$\exists$ a Hilbert space unitarily equivalent to the covariant representation for which $\nu$ is equivalent to a multiple of $M$. $\boxempty$

}\small the graded Hecke algebra\cite{ref03}, \cite{ref19} \cite{ref20}.\normalsize
\vspace{-12pt}
\section{Connections types for Vinberg's $\theta$ groups and period-grading for Vinberg groups\label{conntyp}} Let $G$ be a  simple complex group of adjoint type. {\bfseries {\slshape Def.}}: $\forall$ $\theta$-group ($G_0$, $g_1$) and a vector $X$ $\subset$ its element $g_1$ 
   there $\exists$ a flat G-covariant-derivative $\nabla$ {\itshape locally-monodromic} $connection$ \cite{ref22} $\Rightarrow$  the operator $\nabla^X$ is on the trivial bundle \cite{ref22}: for $m\neq 0$ in the (Kurosh) decomposition Footnote 3, it is possible to choose $m\in M_0$, with M($\lambda$) both a T (V )-module and a $g(V )$ element. \vspace{-14pt}
\subsubsection{Period grading for Vinberg groups }
A Vinberg group can be examined as a $\theta$ group after imposition\footnote{For the sake of the definition of the following von Neumann algebra(s), a few remarks are in order\\{\it Periodically graded semisimple Lie algebras} or, equivalently, {\it periodic gradings of} $g$ (element of a free-composition, also, Kurosh, group) and $\theta$, being $\theta$ the corresponding $m-th$ order of the automorphism, define the \textbf{little Weyl
group} $W(c, \theta)$ in $\mathfrak{c}$: {\bfseries {\slshape Def.}}: the little Weyl group of a graded Lie algebra
 is a subspace c $\in g_1$ and a finite reflection group $W(c, \theta) \in c$;
  for a given $\theta \in \ \  Aut (g) (\theta m = idg)$, an sl2-triple $ \left\{ e, h, f \right\}$ is $\theta$-\itshape{adapted}\upshape $\ \ $, if $\theta(e) = \zeta e$, $\theta(h) = h$, $\theta(f) = \zeta^{−1}f$. $\boxempty$ \\
The {\it regularity of the grading(s)} is classified as: {\bfseries {\slshape Def.}}: the grading is \itshape{\textbf{N-regular}}\upshape $\ \ $ if $g1$ contains a regular nihilpotent element of g; {\bfseries {\slshape Def.}}: the grading is \itshape{\textbf{S-regular}}\upshape $\ \ $ if g1 contains a regular semisimple element of $g$.\\ 
{\bfseries {\slshape Def.}}: the grading is {\it locally free} if $\exists$ $x \in g1$ s. t. $z(x) \cap g0 = \{\ 0 \}\ $; 
it is a centraliser of $\zeta(x) \in g$ if there exists $x \in g_1$ s.t. $\zeta(x)0 = \{\ 0 \}\ $.\\
Let $G$ be a finite unitary reflection group on a complex vector space $\vec{V}$ and $g \in G$ s.t. the $\tilde{\gamma}$-eigenspace, that is $\mathcal{E}$ of($\equiv\in$) $\mathcal{G}$; {\bfseries {\slshape Def.}}: $g$ is {\it maximal} among the $\tilde{\gamma}$-eigenspaces of the  elements of $G$, $\mathcal{G}(\tilde{\gamma})$; {\bfseries {\slshape Def.}}: the centralizer $Z(g)$ of $g$ in $G$ is a stabilizer for the eigenspace $\mathcal{E}$ of $\mathcal{G}$.\\ 
The same terminology applies also to the element $x$ of a $\theta$ group, $\theta1$, $\theta2$: if the corresponding periodic (also, Kurosh, composition) gradings are {\it S-regular} and {\it locally-free}
$\Rightarrow$ $\theta1$, $\theta2$ are \textbf{conjugated by means of an element of the normal complete intersection} $\pi^{-1}\pi \subset g (Int \ \ g)$.\vspace{4pt}\\
As the closable involutive operator(s) of the von Neumann algebra involves non-trivial implications,
further invariants can be defined: {\bfseries {\slshape Def.}}: the $\eta-$ \textbf{invariants} are defined as \cite{ref20}
\begin{equation}\label{pareta}
\eta_{\pm}=\lim_{t\rightarrow 0_{\pm}} \eta, \ \  \eta(\nabla_i)=\pm(*\nabla_i-\nabla_i*),\ \ \eta_+=\eta_0+\sum_{i\ge 1}\sigma_i, \ \ \eta_-=\eta_0+\sum_{i\ge 1}(-1)^i\sigma_i 
\end{equation}
the domains of the arithmetic groups, as well as those of the non-arithmetical ones, are delimited by surfaces of equal probabilities (as (integral) functions of the composition of two invariants).} of period grading, as implying by Kostant-Weierstrass sections (see Appendix (\ref{cbm})) for $g_0$ {\it semisimple} and $m>3$:
with $\theta \in Aut $ any automorphism of order $m$, if $\theta$ is \textbf{S-regular} and \textbf{locally free}, then the corresponding $\theta$ -group admits
a KW-section.\\
If $\theta$ is \itshape{not locally free}\upshape $\ \ $, \textbf{only one of the two subgroups}, say $\theta_1$, \textbf{is admitted for the KW section}.\\
 If both $\theta1$ and $\theta2$ are \itshape{S-regular but
not locally free}\upshape $\ \ $, \textbf{they are not conjugate}.\\
If $\theta \in Aut \ \  g$ is \itshape{N-regular}\upshape, $\exists$ two automorphisms given y the suitable restrictions.\\
If $\theta$ is \itshape{N-regular}\upshape, its \textbf{ideal} in $\kappa[g]$ is
\textit{generated by the basic invariants}  $\pi^{−1}(\pi(c))$ \textbf{i.e. a complete intersection}.\\
{\it Complex reflections} The Weyl group contains complex transformations, i.e. {\it complex reflections}\footnote{ for reflections, the definition of the direct sum of
character sheaves as a character sheaf follows
\cite{ref03}, the algebra of invariants of any of its $\theta$-group is free \cite{ref34}, \cite{ref03}.
}.



\begin{figure}[htbp]
\begin{minipage}{15pc}
\includegraphics[width=15pc]{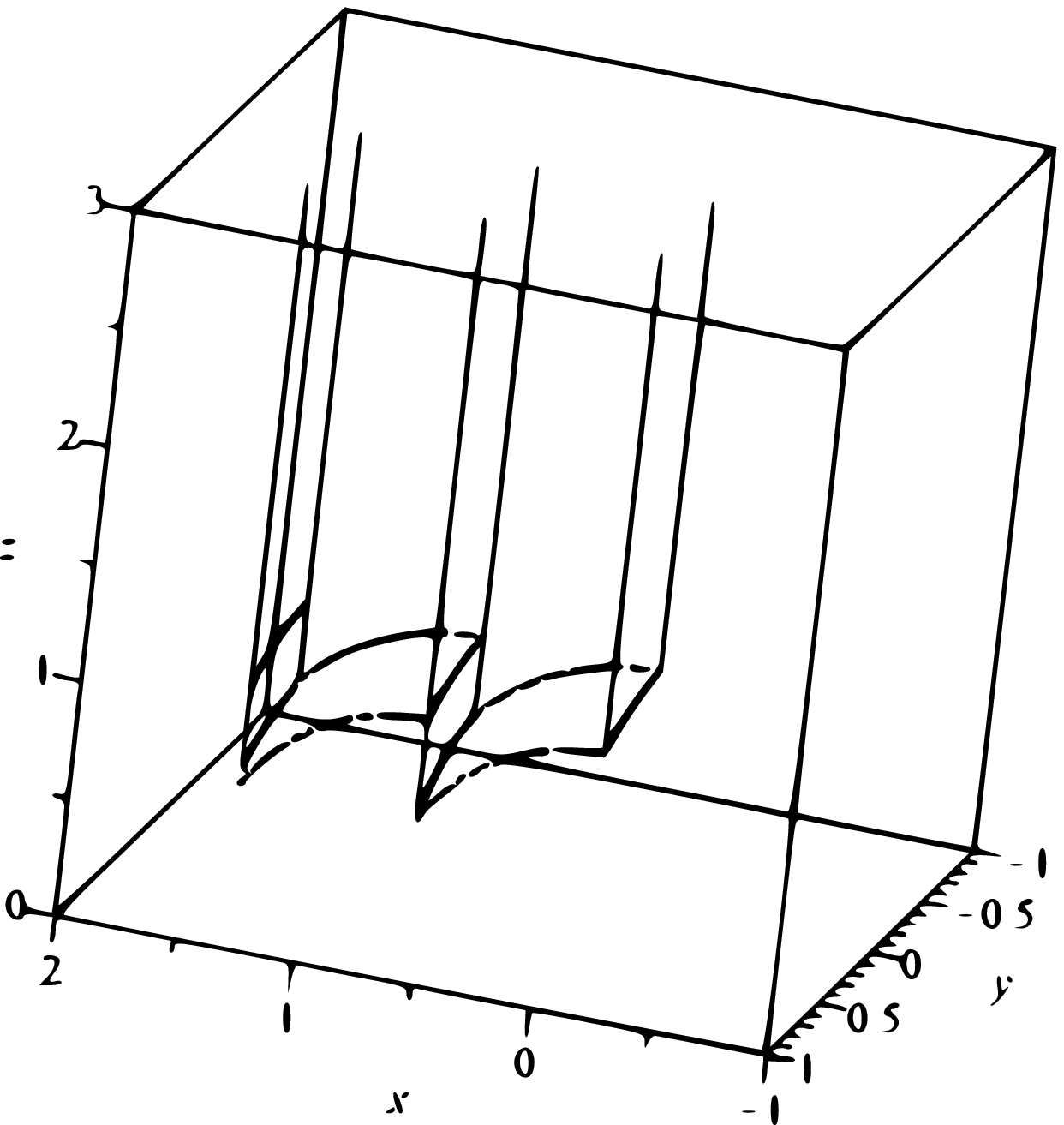}
\caption{\label{labelfigure7} \small $n=1$ {\bf grading for the Vinberg group} $x_\beta=\frac{\sqrt{5}}{3}$. \scriptsize The domain of the subgrouppal structure for the $n=1$ grading of the Vinberg (non-arithmetical) group characterized by\\
$\bullet \ \ $  $\beta=\frac{4}{5}\pi$, i.e. $x_\beta\simeq \frac{\sqrt{5}}{3}$;\\
$\bullet \ \ $$0\leq x\left(\Gamma_0\left(x_\beta=\frac{\sqrt{5}}{3}\right)\right)\leq \frac{\sqrt{5}}{3}$, $0\leq y\leq \frac{1}{2}$;\\
the complex at $x=\tilde{m}\frac{\sqrt{5}}{3}$, $\tilde{m}\in \mathbb(Z)$ is sketched. \normalsize}
\end{minipage}\hspace{2pc}%
\begin{minipage}{15pc}
\includegraphics[width=15pc]{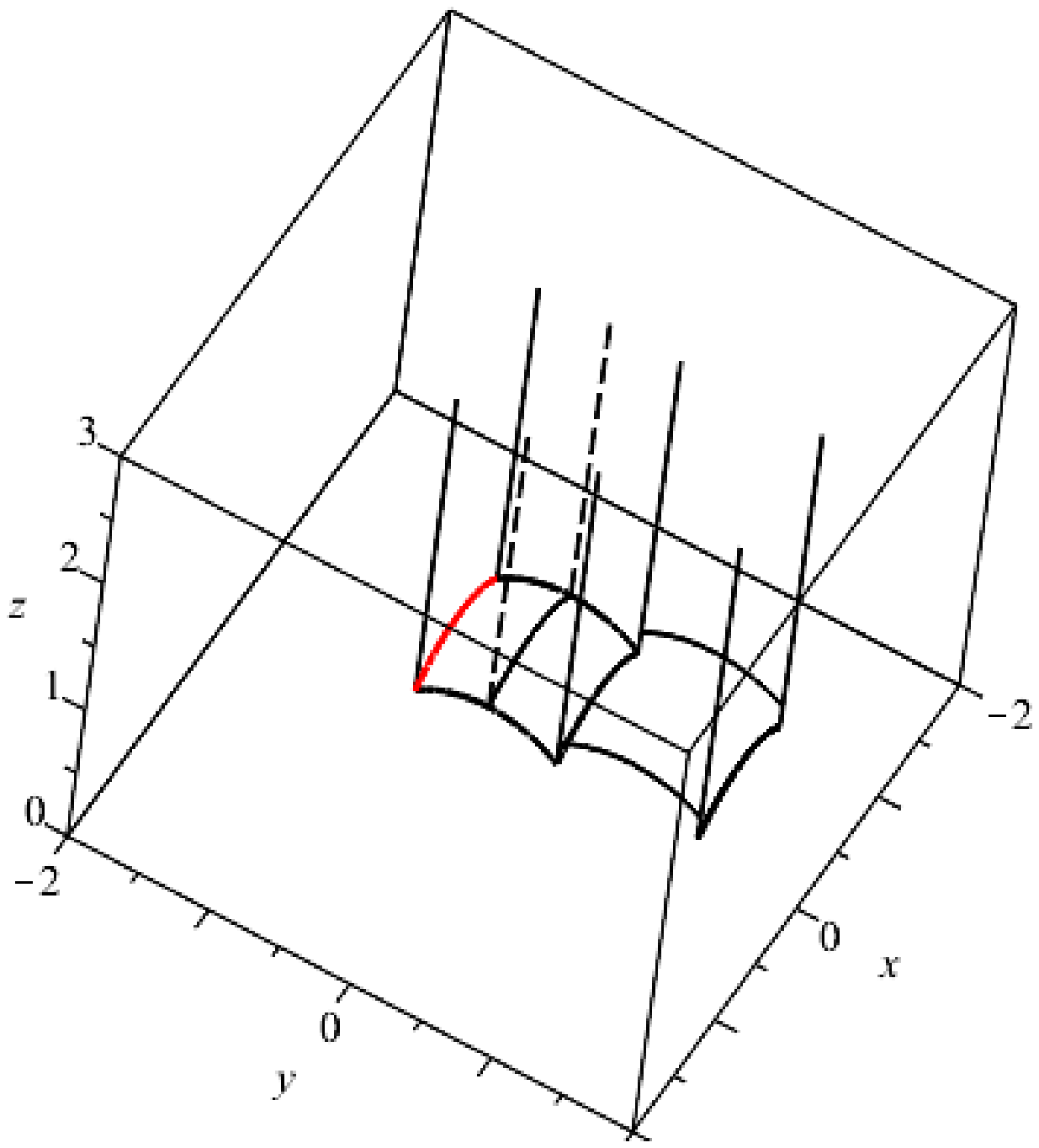}
\caption{\label{labelfigure8} \small $n=1$ {  \bf Grading for the  subgrouppal structure}    $\Gamma_0\left(Vinberg \left( x_\beta>\frac{1}{2}\right)\right)$. \scriptsize The domain of the subgrouppal structure for the $n=1$ grading of the subgrouppal structure $\Gamma_0\left(Vinberg \left(x_\beta>\frac{1}{2}\right)\right)$;\\
the simplex $v=0\pm\frac{m'}{2}$ is sketched, $m'=2m+1$.\normalsize}
\end{minipage} 
\end{figure}
\vspace{-9pt}

 \begin{figure}[htbp]
\begin{minipage}{15pc}
\includegraphics[width=15pc]{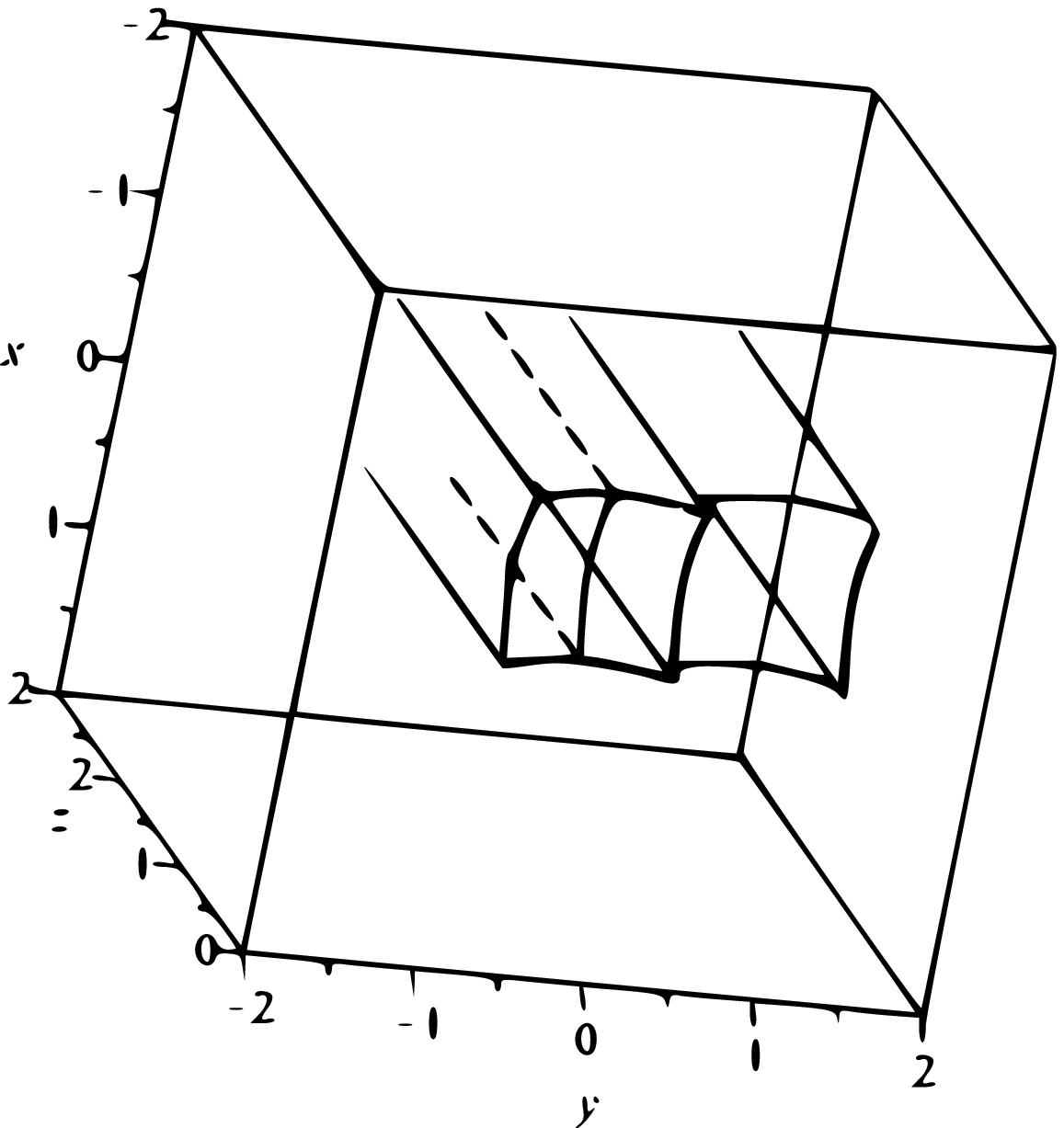}
\caption{\label{label} {\bf The  subgrouppal structure}    $\Gamma_0(Pic)$. \scriptsize The domain of the subgrouppal structure $\Gamma_0(Pic)$;\\
the simplex $u_1=0\pm\frac{m'}{2}$ is sketched, $m'=2m+1$. \normalsize}
\end{minipage}\hspace{2pc}%
\begin{minipage}{15pc}
\includegraphics[width=15pc]{gammazerovinbergottuso.eps}
\caption{\label{labelfigure50} \small {\bf The subgrouppal structure}   $\ \ $$\Gamma_0\left(x_\beta=\frac{\sqrt{5}}{3}\right)$ $\ \ $ {\bf of the Vinberg group}. \scriptsize The domain of the  subgrouppal structure for the Vinberg (non-arithmetical) group characterized by\\
$\bullet \ \ $ an angle $\beta=\frac{4}{5}\frac{\pi}{2}$, i.e. $x_\beta\simeq \frac{\sqrt{5}}{3}$;\\
$\bullet \ \ $$-\frac{\sqrt{5}}{3}\leq x\left(\Gamma_0\left(x_\beta=\frac{\sqrt{5}}{3}\right)\right)\leq \frac{\sqrt{5}}{3}$, $0\leq y\leq \frac{1}{2}$;\\
the simplex at $x=\tilde{m}'\frac{\sqrt{5}}{3}$, $\tilde{m}'\in{\mathbb Z}$ is sketched. \normalsize}
\end{minipage} 
\end{figure}

\normalsize
\vspace{-9pt}
\section{Outlook: Hamiltonian analysis, Results and Remarks}
In the Hamiltonian analysis, {\it cuspidal datum} and {\textit initial cuspidal data} for {\textbf non-trivial remainders and virtually surjective pairs on subgroups} can be compared, i.e. {\bf also in the case of complexes}. The existence of the invariants (\ref{pareta}) allows one for the analysis of initial conditions and boundary conditions.
\vspace{-14pt} 
\subsubsection{Particles representations and eigenfunctions}
In the orthonormal directions of the the embedded space the target map  $\rho$ ensures a Wigner-Bargman representation via the $\eta$-invariants (\ref{pareta}) as
\small
\begin{equation}
\eta^{\mathcal{N}}\alpha(ra)=\eta^{\mathcal{N}}\rho(r)\alpha(a)
\end{equation}
\normalsize
for any $\mathcal{N}$-th iteration of the invariants on $r\in \Lambda$, with $\Lambda$ Poincar\'e transformations:
 any Hamiltonian system implying a continuous Hamiltonian flow must contain at least a second order differential operator in the {\itshape-}$ $complex \ \ $v\ \ $ direction (i.e. in the direction along which the grading is not considered).\\ 


\begin{figure}[htbp]
\includegraphics[width=14pc]{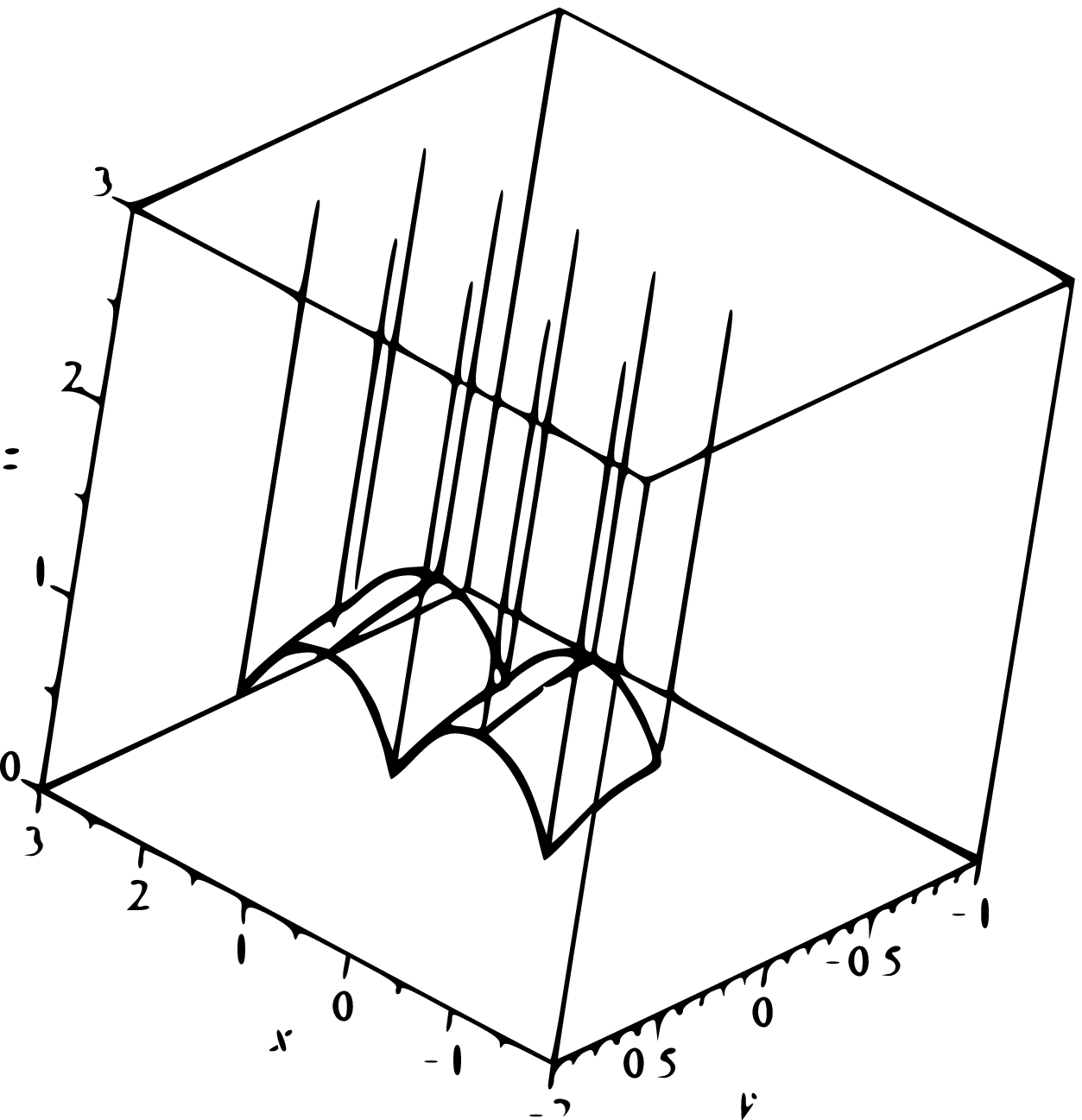}\hspace{2pc}%
\begin{minipage}[b]{28pc}\caption{\label{labelfigure51} {\small \bf grading for the  subgrouppal structure}    $n=1 \ \ \left(Vinberg \left(x_\beta>\frac{1}{2}\right)\right)$. \scriptsize The domain of the subgrouppal structure for the grading of the subgrouppal structure $\Gamma_0\left(Vinberg: x_\beta>\frac{1}{2}\right))$;\\
the complexes $x=0\pm 2x_\beta$ are sketched.\vspace{4pt}\\
\tiny
The tangent bundle on a delimiter of the complex coincides with that of the elsewhere smooth manifold
 (containing the complex) on the delimiter of the complex itself, and the tangent fibration for a simplex which does not include a cuspidal point does not imply discontinuities on (invariants) for (at least) holomorphic functions; such definitions for meromorphic (eigen-)functions are well defined
\cite{ref39}:$i)$ initial cuspidal data on {\bf complexes} and on initial data on subspaces that contain {\bf non-only simplices}
are non-compatible with virtually surjerctive pairs, i.e. non-compatible with local interaction(s);
$ii)$ {\it initial cuspidal data on} {\bf complexes} and on {\it initial data on subspaces that contain {\bf non-only simplices}} is suited for the grading of modular structures (for desymetrized domains); $i_a)$ for initial cuspidal data implying non-trivial remainders (signatures) in the sum of all the comprehended $\eta$ invariants if non-symmetric parts of the domains are considered, convergence of the reminders can be looked for even for strongly non-local (or. i.e. non-contact) interaction, which case is still suited for spin-like interaction (i.e. {\it lattice models})\cite{ref45}; under suitable hypotheses, the representation of scalar can be assured only locally  on Minkowski spacetime \cite{ref86}; $\it iii)$ {\bf in the direction orthonormal to the complex(ified) variable(s)} for $\Gamma$ subgroups and $\Theta$ ones of non-arithmetical groups with boundary conditions and initial cuspidal data on {\it simplices}, $\eta-$invariants do not have reminders; $iii_b)$ the grading of modular structures 
for the Hamiltonian flow (on the Hamiltonian coordinate variable of the space in which the structures are embedded)  {\bf not including simplices} among {\it at least} at least within the  $imaginary$ ($v$) direction 
 has to be compared with modular structures which might be also not be symmetric.
 \normalsize}
\end{minipage}
\end{figure}

  The presence of simplices and complexes does not imply interaction \cite{ref37}, \cite{ref39} also as from the\footnote{Be $\mathcal{S}$ the corresponding closable involutive operator of the von Neumann algebra. The Fock space is spanned by the operators
$\Xi\equiv$ CPT, for which the composition $ \mathcal{S}\equiv\Xi\Xi_0$ holds, with $\Xi_0$ modular conjugation; here,
eigenfunctions (physical states) $K_{{\mathcal F}}$ of operators which are CPT-invariant $2\pi$ rotations  (after the definition of Hecke algebras of type $D$) (\ref{app11}) 
\small
\begin{equation}
K_{{\mathcal F}}=\sum_ne^{i\pi sn^2}P_N,
\end{equation}
\normalsize
}
\footnote{
, with, the spin-statistics-theorem, either $s\in\mathbb{N}$, or $s=j/2$ $,\in\mathbb{N}$, 
as summarized in \cite{ref27}. (At least) holomorphic functions are eigenfunctions both of the Hamiltonian and of the $\eta-$ invariant (\ref{pareta}); after grading, the symmetry group is still a free diffeomorphism group.\\
The following cases are distinguished: $\bullet$ $\mathcal{S}=1$ trivial; $\bullet$ $\mathcal{S}\neq1$ non-trivial restrictions: $\bullet \bullet$ interaction or $\bullet \bullet$ issues concerning the manifolds and/or (Haar) measure \cite{ref6}} Fock-representation-space $K_{{\mathcal F}}$.

The Hamiltonian formalism proved one with the simplest invariant structure (\ref{pareta}):
${\bf i)}$ $\forall$ Poincar\'e complex $\exists$$!$ the tangent fibration up to equivalence classes (i.e. of fibrewise homotopy);\\
${\bf ii)}$ $\forall$ Poincar\'e (-complex) embedding, the normal fibration is the tangent fibration of the Poincar\'e complex \cite{ref38}
(the tangent fibration of a smooth manifold is an invariant of such a manifold up to the fibration equivalence):
$\Rightarrow$ there is one and only one class of pairs that trivialize the neightborhood (of a point on the complex) \cite{ref28}. Indeed, a manifold $V$ can be foliated covariantly to a \itshape holonomy gruppoid \upshape $\ \ $ $\mathcal{G}$ on which can apply Lie groups, with an induced map $H^*(V)=H^*(\mathcal{BG})$: the grading of modular structures (for desymetrized group domains) provides a measure for the Hamiltonian flow.The analysis of initial-values-conditions for the evaluation of the invariants includes non-trivial results.\vspace{4pt}\\
Starting from the analyses \cite{ref8} and \cite{ref42}, in \cite{ref41}, the case  elliptic modular equations, which applies also for Picard-Fuchs equations, are studied to generate elliptic curves of $genus-zero$ congruence subgroup(s) \cite{ref43} of $\Gamma_0(N)$, which admit a covering map $X_0({\mathbb N})$  to $X(1)$, and associated modular forms vanishing only at an {\it infinite} simplices or at an {\it infinite} complexes. The examinations can be pursued the analysis of \cite{ref44} for curves of $genus-one$. This analysis is subsequent to the fact that the pertinent equations and the fibration of the pertinent manifolds are not from a Banach space \cite{SCRIVI}, but are {\it at least} on Hecke pencils \cite{SCRIVI2}.
\vspace{-13pt}
\small
\section*{Acknowledgments}
OML is grateful to Comenius University in Bratislava,
Faculty of Mathematics, Physics and Informatics,
Department of Theoretical Physics and Physics Education- KTFDF for warmest hospitality. The work of OML was partially supported by the SAIA- NS'P (Slovak Academic Information Agency- National Stipendium Programme) Scholarship for incoming International Visiting Researchers, and partially by the KTFDF Conference Grant $O-06-107/0014-00$. OML would like to thank Prof. Vladim$\acute{i}$r Balek and Prof. Peter Pre$\check{s}$najder for reading the proceeding.  
\normalsize
\vspace{-9pt}
\begin{figure}[htbp]
\begin{minipage}{15pc}
\includegraphics[width=15pc]{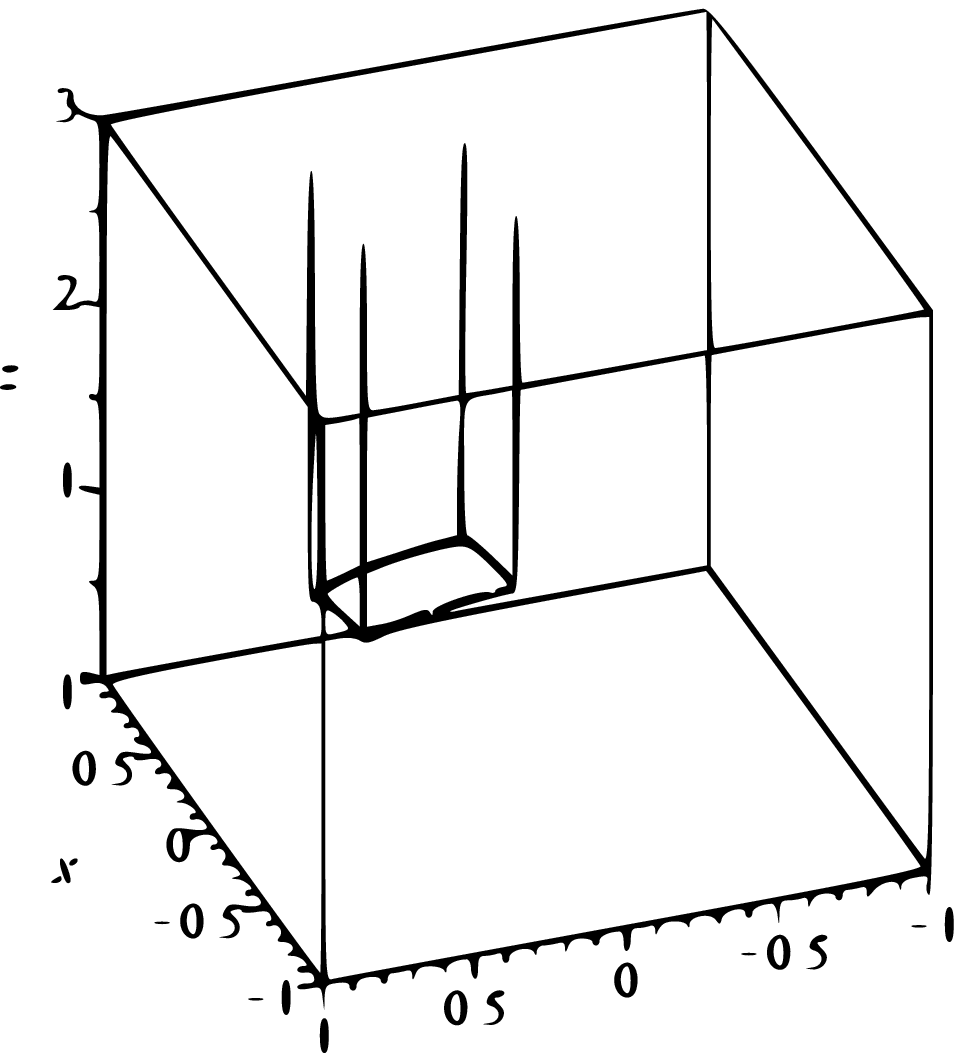}
\caption{\label{labelfigure1}{\small \bf The desymmetrized Picard group.} \scriptsize The domain of the (arithmetical) Picard group, i.e. $\ \ $$x_\beta=\frac{1}{2}$ \normalsize.}
\end{minipage}\hspace{2pc}%
\begin{minipage}{15pc}
\includegraphics[width=15pc]{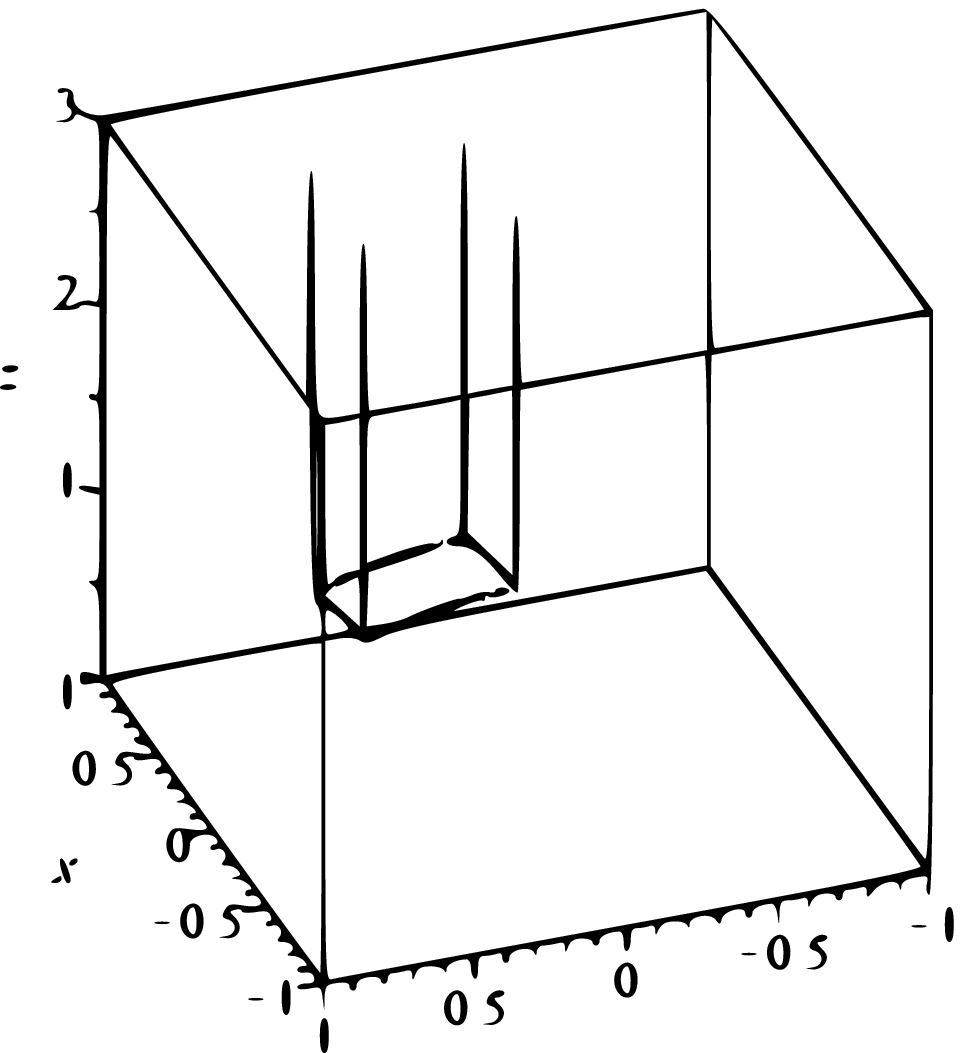}
\caption{\label{labelfigure2}{\small \bf The desymmetrized Vinberg group}. \scriptsize The domain of a (non-arithmetical) Vinberg group, for which $x_\beta=\frac{\sqrt{2}}{3}$;\\
$0\leq x(x_\beta)\leq \frac{\sqrt{2}}{3}$, $0\leq y\leq \frac{1}{2}$;\\
$\bullet \ \ $ with $\vec{a}\in \mathbb R^n$, i.e. $a=(1, 0)$, and the trivial automorphism $A=1$ wrt the $x$ direction.\normalsize}
\end{minipage} 
\end{figure}
\appendix
\small
\section{A description of the fiber}
Kostant-Weierstrass sections are a  linear subvariety $\tilde{v}$ which induces an isomorphism of affine varieties. There exists a $grading $ \cite{ref29}
 $g = g(0)\otimes g(1)$ of the the symmetric space
decomposition for $m=2$.
\vspace{-14pt}
\subsubsection{Subalgebra isomorphisms and canonical basis elements\label{cbm}}
Let $H$ be  a unital associative algebra isomorphic to $H_A$. $H$ restricts to an isomorphism between the A subalgebras
of $H$, and $H_A$ generated by $f(n)$: $F\cdot \in \mathcal{H}$ is a {\it unique mixed structure} such that the $\mathcal{H}$ elements in 
 $H_C$ provide with the identification $U^A$ both $\subset$ $\mathcal{H*}$ and $\subset$ $H_A$: this is a 
{\it subalgebra isomorphisms}.\\
{\it Canonical basis elements} are given as a mixed structure $\Rightarrow$ the product of canonical basis elements $b_1\cdot b_2$ expanded  into a linear combination of canonical basis elements $\sum c^{b_3}_{b_1b_2}$ of $(v)$ in $\mathcal{H}$ corresponds to 
$v$ under (for) the subalgebra isomorphisms identification, as in \cite{ref23}:
be A a {\it finite dimensional K-algebra} and  A is of infinite representation type, with $I$ is a two–sided ideal of A, in which case,
$A/I$ has infinite representation type: $\Rightarrow$ A is of infinite representation type; be B a direct summand of A as a (B,B)$-$bimodule  $\Rightarrow$ if B is of infinite representation type, then so is A $\Leftrightarrow$ so is B.  $\boxempty$ \\
 As in \cite{ref48}, be A be a {\it finite dimensional K}$-${\it algebra}, and be $P_1, . . . P_l$  the complete set of projective indecomposable A$-$modules, up to isomorphism: $i)$ if $End\ \ A (P_j)$ has infinite representation type for some $1\le j\le l$, then A is of infinite
representation type;
$ii)$ for each $1\le j\le l$, the algebra $End\ \ A (P_j$) has finite representation type iff $End\ \ A (P_j)\simeq
K[x]/<x^m>$ for some integer $m\ge0$ (which depends on $j$), $m\equiv m(j)$, $m\in {\mathbb N} /0$ $\boxempty$ ;
for a A$-$ $module$ $M$, under suitable assumptions, for a  finite-dimensional K-algebra there might exist  $P_{\mathfrak{I}1},...,  P_{\mathfrak{I}_l}$ a set of projective indecomposable A-modules\footnote{According to the Dipper-James criterion, there is a direct summand in the $q$ permutation $Y^\lambda$ s.t. $Y^\lambda$ contains a unique submodule $S^\lambda$: {\bfseries {\slshape Def.}}: the {\it fundamental domain for the action of its indecomposable groups} is the closure of the affine Weyl group;  
for $\mathcal{F}_{\mathcal{M}}$ the boundary of a manifold $\mathcal{M}$ and $\mathcal{V}$ a {\it direct sum of indecomposable modules} with {\textbf vertices (or, also, vertex)} not conjugated on $\mathcal{F}_{\mathcal{M}}$; by $m_i$, {\it any indecomposable summand is the direct sum of the indecomposable summands} with the $S^{\lambda}$ pullback on the direct $\rho$ map.}, up to isomorphism $\boxempty$;
for $m\ge2$, and
G a complex reductive group, whose elements
g = Lie(G) are generated by the algebra Lie(G(0)) = g(0), with
 G(0) the normalizer for g(1) $\Rightarrow$ {\bfseries {\slshape Def.}}: $d\theta$ is the automorphism of order m of G.
\vspace{-12pt}    
\subsection{Induced gradings}For the $\theta$-induced grading\footnote{
The little Weyl group $Wc$ is generated by pseudoreflections, and
the associated isomorphism $\kappa$ is a polynomial ring.}, 
a Kostant-Weierstrass (KW)-section for the pair (G, $\theta$) is a linear subvariety $v$ of g(1)
for which the restriction of functions $k[g(1)]G(0) \rightarrow \kappa[\tilde{v}]$ is an isomorphism.\\   
The KW-section $\exists$ if g(0) is semisimple, for N-regular $\theta$,
 for all classical graded Lie algebras in zero or good positive
characteristic p, p not dividing m; from \cite{ref33}, for $p > 3$ and $\theta$ an outer automorphism of G, KW-sections exist {\itshape for all classical graded Lie algebras in zero or odd positive characteristic} $\boxempty $.
\vspace{-12pt}
\subsubsection{Hecke algebra of the type D\label{app11}} A Hecke algebra of the type D is defined as {\bfseries {\slshape Def.}}: orthogonal group s.t. $i_a)$ needed classification  of the isomorphism classes of the indecomposable modules; $i_b)$ admits a canonical basis; $i_c)$ for $D^\lambda\neq0$, is defined as {\it disjoint union} of $\mathcal{KBP}$(n); $i_d)$ for $S^\lambda: D^\mu$  the canonical basis in a combinatorial canonical basis, which plays the role of a Fock (occupation number);
$i_e) $the characteristic of F is odd \cite{ref26}. $\boxempty$
\vspace{-12pt}
\subsubsection{(KW)-Section for the Vinberg group.} The section $W_1$ for the subgroup $\theta_1$, $W_1\equiv W_1(\theta_1)$ ( not necessarily the same as 
$W^\sigma$ or $W(c, \theta)$, a reflection group in c, 
with $\sigma=\theta\mid_i$) defined only if
$m_j$ is a co-exponent of $W(c, \theta)$, s.t. the identical element is $\epsilon_1\zeta^{d_1}=I$,i.e composed by the stabilizer $\zeta(x) \in g$ s.t. $\epsilon_i$ $(i = 1, . . . , l)$, which depend only on
the connected component(s) of $Aut \ \  g$ containing $\theta$, s.t.  $\kappa_i = \kappa_i(\theta,m)$, $\forall$ $(i \in\mathbb{Z}_m)$).\\
$\Rightarrow$  $\theta \in Aut \ \  g$ is N-regular; the triple ($ \left\{ e, h, f \right\} $) is a $\theta$-adapted regular
sl2-triple. $\boxempty$\\
The following cases distinguish for $m_j$ an exponent of $W(c, \theta)$ iff $\epsilon_j\zeta^{m_j}=\zeta^{-1}$:   
 if $\theta$ is also S-regular, $m_j$ is a co-exponent of $W(c, \theta)$ iff $\epsilon_j\zeta^{m_j}=\zeta$;    
if $\theta2$ is not N-regular, it \textit{might} happen that $\exists$ a finite reflection group (multiple of $m$) because $\theta2$ is S-regular; if $\theta$ is neither N-regular nor S-regular, still $dim\ \  g1\ \  G0 = k_{−1}$.
\vspace{-5pt}
 \begin{figure}[htbp]
\begin{minipage}{15pc}
\includegraphics[width=15pc]{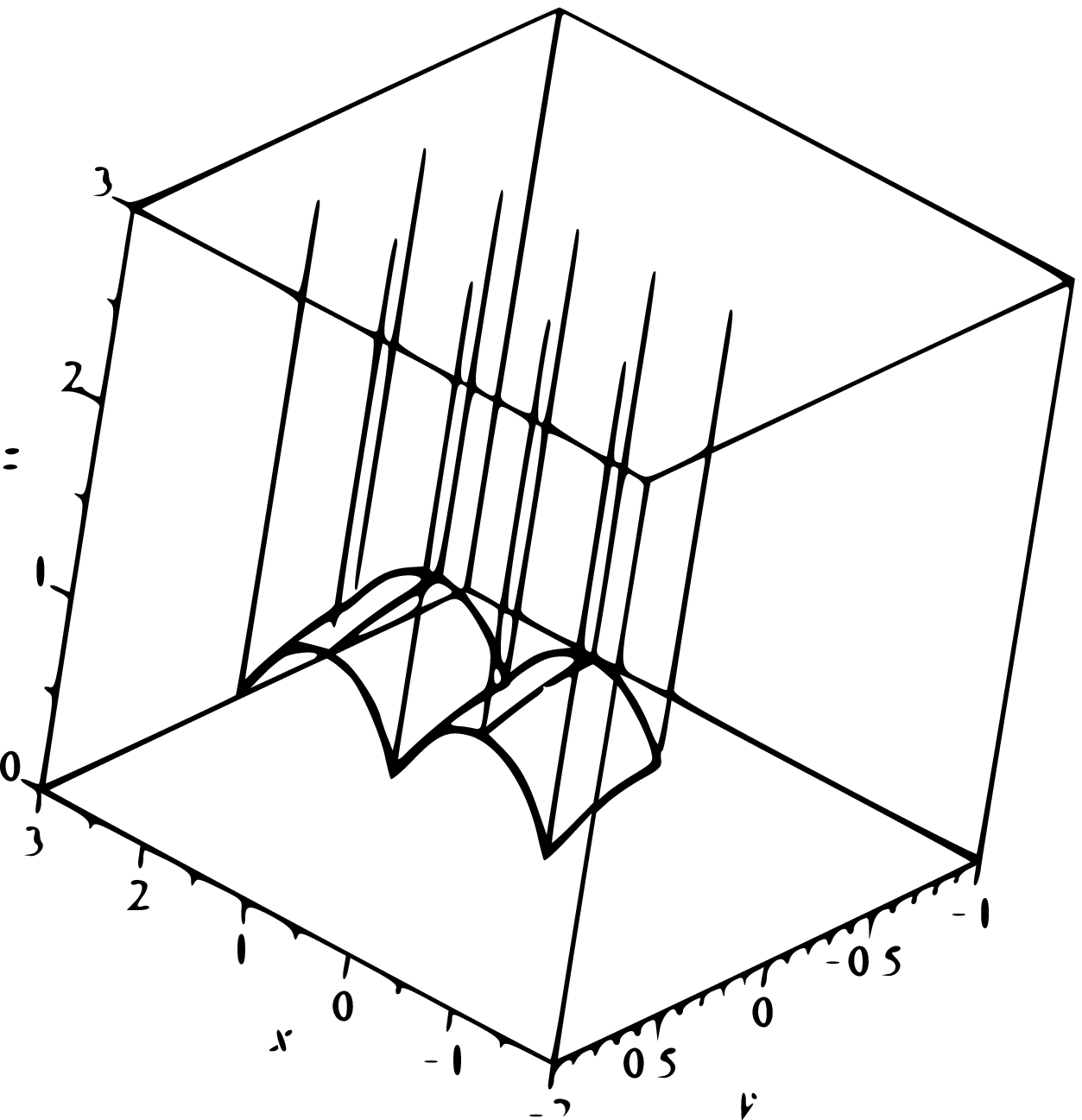}
\caption{\label{labelfigure20}\small  {\bf Grading for the  subgrouppal structure}    $n=1 \ \ (Vinberg- x_\beta>\frac{1}{2})$. \scriptsize The domain of the subgrouppal structure for the grading of the subgrouppal structure $\Gamma_0(Vinberg- x_\beta>\frac{1}{2})$;\\
the complex $u=0\pm 2x_\beta$ is sketched. \normalsize}
\end{minipage}\hspace{2pc}%
\begin{minipage}{15pc}
\includegraphics[width=15pc]{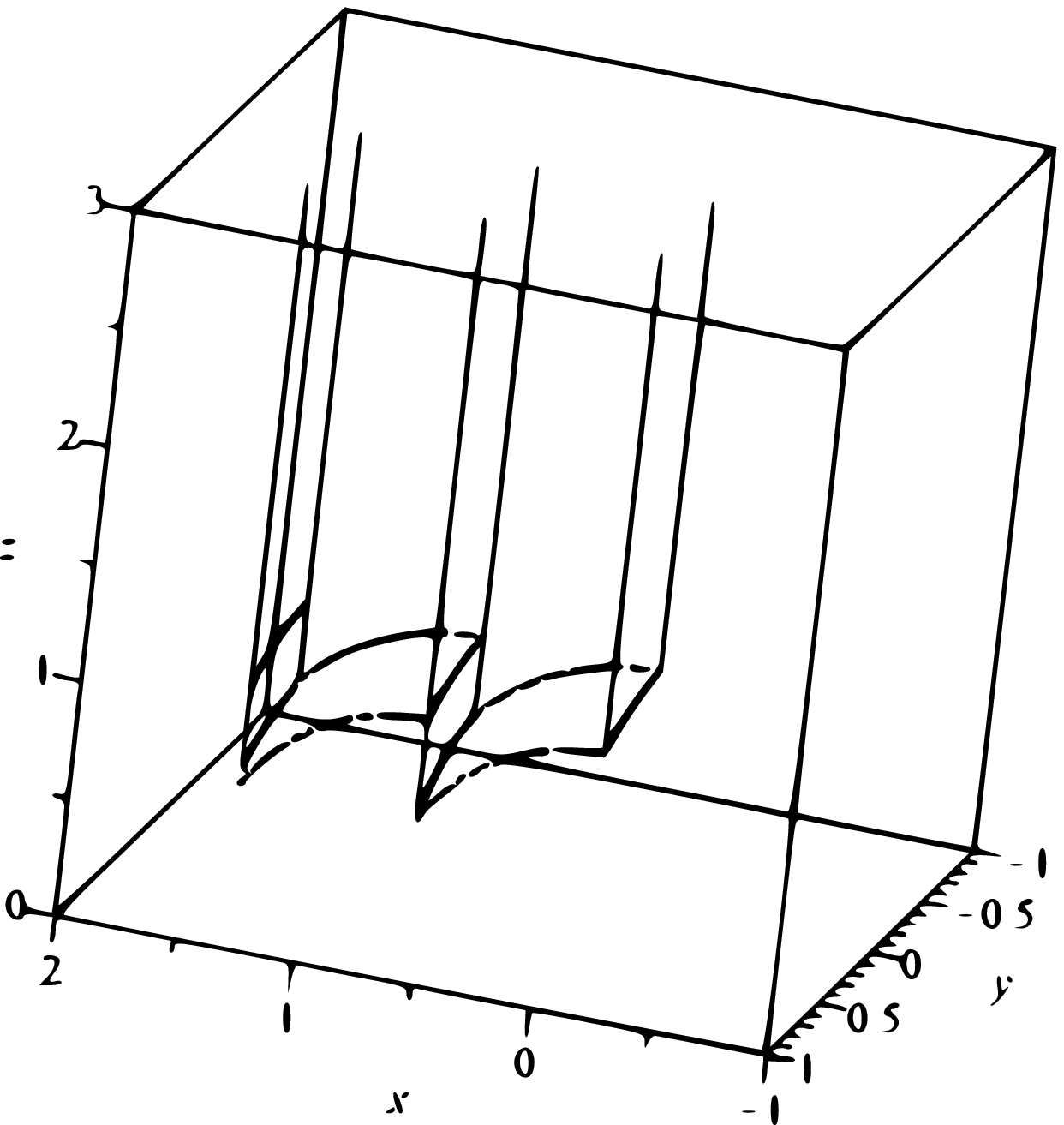}
\caption{\label{labelfigure21} \small $n=1$ {\bf Grading for the Vinberg group} $x_\beta=\tfrac{\sqrt{5}}{3}$. \scriptsize The domain of the subgrouppal structure for the $n=1$ grading of the Vinberg (non-arithmetical) group characterized by\\
$\bullet \ \ $  $\beta=\frac{4}{5}\pi$, i.e. $x_\beta\simeq \frac{\sqrt{5}}{3}$,\\
$\bullet \ \ $$0\leq x\left(\Gamma_0\left(x_\beta=\frac{\sqrt{5}}{3}\right)\right)\leq \frac{\sqrt{5}}{3}$, $0\leq y\leq \frac{1}{2}$;\\
   
the complex at $x=\tilde{m}\frac{\sqrt{5}}{3}$, $\tilde{m}\in \mathbb(Z)$ is sketched. \normalsize}
\end{minipage} 
\end{figure}

\end{document}